\documentclass[10pt,twocolumn,twoside, journal]{IEEEtran}

\usepackage{cite,amsmath}
\usepackage{amssymb, bbm, bm, dsfont, url}
\usepackage{relsize} % for \smaller to work.=

\usepackage{pifont, url}% http://ctan.org/pkg/pifont
\ifCLASSINFOpdf
   \usepackage[pdftex]{graphicx}
   \DeclareGraphicsExtensions{.pdf,.jpeg,.png}
\else
   \usepackage[dvips]{graphicx}
   \DeclareGraphicsExtensions{.eps}
\fi

\begin{document}

\title{Streaming End-to-End Multi-Talker Speech Recognition}
%\name{Liang Lu, Naoyuki Kanda, Jinyu Li and Yifan Gong}
%\address{Microsoft Corp., USA}
%\email{\{liang.lu, naoyuki.kanda, jinyli, yifan.gong\}@microsoft.com}

%\begin{document}

\author{Liang~Lu,~\IEEEmembership{Member,~IEEE},  Naoyuki Kanda,~\IEEEmembership{Member,~IEEE}, Jinyu Li~\IEEEmembership{Member,~IEEE} and Yifan~Gong,~\IEEEmembership{Fellow,~IEEE}
\footnote{Copyright (c) 2021 IEEE. Personal use of this material is permitted. However, permission to use this material for any other purposes must be obtained from the IEEE by sending a request to pubs-permissions@ieee.org.}
%\thanks{Manuscript received -; revised -}
\thanks{The authors are with Microsoft Corp., USA; Copyright (c) 2021 IEEE.%email: {\smaller \tt {\{liang.lu, naoyuki.kanda, jinyli, yifan.gong\}@microsoft.com}}

%Copyright (c) 2021 IEEE. Personal use of this material is permitted. However, permission to use this material for any other purposes must be obtained from the IEEE by sending a request to pubs-permissions@ieee.org.
}%
}

%\markboth{IEEE Signal Processing Letters,~Vol.~X, No.~X,~2021}%
{}

\maketitle
\begin{abstract}
End-to-end multi-talker speech recognition is an emerging research trend in the speech community due to its vast potential in applications such as conversation and meeting transcriptions. To the best of our knowledge, all existing research works are constrained in the offline scenario. In this work, we propose the Streaming Unmixing and Recognition Transducer (SURT) for end-to-end multi-talker speech recognition. Our model employs the Recurrent Neural Network Transducer (RNN-T) as the backbone that can meet various latency constraints. We study two different model architectures that are based on a speaker-differentiator encoder and a mask encoder respectively. To train this model, we investigate the widely used Permutation Invariant Training (PIT) approach and the Heuristic Error Assignment Training (HEAT) approach. Based on experiments on the publicly available LibriSpeechMix dataset, we show that HEAT can achieve better accuracy compared with PIT, and the SURT model with 150 milliseconds algorithmic latency constraint compares favorably with the offline sequence-to-sequence based baseline model in terms of accuracy.
\end{abstract}
%\noindent\textbf{Index Terms}: Overlapped speech recognition, Streaming, Unmixing transducer,  Heuristic error assignment training

\ifCLASSOPTIONpeerreview
\begin{center} \bfseries EDICS Category: SPE-RECO \end{center}
\else
\fi
\IEEEpeerreviewmaketitle

\section{Introduction}

\IEEEPARstart{O}{verlapped} speech is ubiquitous among natural conversations and meetings. For automatic speech recognition (ASR), recognizing overlapped speech has been a long-standing problem. A common practice is to follow the divide-and-conquer strategy, e.g., applying speech separation cascaded with a single-speaker speech recognition model~\cite{chen2020continuous}. While this approach has enjoyed significant progress thanks to the achievement in deep learning based speech separation~\cite{yu2017permutation, hershey2016deep, luo2018tasnet}, there are two key drawbacks with this paradigm. Firstly, the overall system is cumbersome, especially given the increasing complexity of both speech separation and speech recognition modules. Consequently, maintaining and developing the cascaded system requires significant engineering effort. Secondly, each module in the cascaded system is optimized independently, which does not guarantee the overall performance improvement.
%For instance, 
% improvement in the speech separation front-end may not result in word error rate reduction of overlapped speech recognition~\cite{von2020multi}. 
%Last but not least, the cascaded system could pose longer latency, and therefore it is not very applicable for the streaming transcription scenario. 

Recently, there have been considerable amount of work on the end-to-end approach for overlapped speech recognition.  End-to-end speech recognition models, such as Connectionist Temporal Classification (CTC)~\cite{graves2006connectionist, miao2015eesen, Li18CTCnoOOV, audhkhasi2018building}, attention-based sequence-to-sequence model (S2S)~\cite{chorowski2015attention, lu2016training, chan2016listen, chiu2018state},  and Recurrent Neural Network Transducer (RNN-T)~\cite{graves2012sequence, he2019streaming, Li2020Developing} have been explored to address this challenge. In particular, Settle et al.~\cite{settle2018end} proposed a model with joint speech separation and recognition training. Chang et al.~\cite{chang2019end} applied multi-task learning with CTC and S2S to train an end-to-end model for overlapped speech recognition. Kanda et al.~\cite{kanda2020serialized} proposed Serialized Output Training (SOT) for S2S-based end-to-end multi-talker speech recognition. RNN-T has also been investigated for overlapped speech recognition in~\cite{tripathi2020end} in an offline setting with bidirectional long short-term memory (LSTM)~\cite{hochreiter1997long} networks and auxiliary masking loss functions. Compared with the joint speech separation and recognition approach using an hybrid model, the end-to-end approach enjoys lower system complexity and high flexibility~\cite{yu2017recognizing, qian2018single}. While the progress in end-to-end overlapped speech recognition is promising, to the best of our knowledge, all previous studies only consider the offline condition, which assumes that the overlapped audio has been segmented. Thus, these systems cannot be deployed for streaming first-pass speech recognition scenarios that require low recognition latency such as online speech transcription for meetings and conversations. %Unfortunately, this is a poor assumption, as the segmentation for overlapped speech itself is a challenging problem. In most speech recognition tasks, speech signal comes in a continuous mode, and it requires the recognizer to be streaming for good user experience. In these scenarios, offline models cannot be deployed. 

In this paper, we propose the Streaming Unmixing and Recognition Transducer (SURT) for multi-talker speech recognition. Our model relies on RNN-T as the backbone, and it can transcribe the overlapped speech into multiple streams of transcriptions simultaneously with very low latency. In this work, we investigate two different network architectures. The first architecture employs a mask encoder to separate the feature representations, while the second model uses a speaker-differentiator encoder~\cite{chang2019end} for this purpose. To train SURT, we study an  approach similar to the one applied in~\cite{tripathi2020end}, which we refer to as Heuristic Error Assignment Training (HEAT) for the clarity of presentation. This approach can be viewed as a simplified version of the widely used Permutation Invariant Training (PIT)~\cite{yu2017permutation} by picking only one label assignment based on heuristic information. Compared with PIT, HEAT consumes much less memory, and is more computationally efficient. %More importantly, it is more scalable to the number of speakers. 
To evaluate the proposed SURT model, we performed experiments using the LibriSpeechMix dataset~\cite{kanda2020serialized}, which simulate the overlapped speech data from the LibriSpeech corpus~\cite{panayotov2015librispeech}. We show that SURT can achieve strong recognition accuracy with 150 milliseconds algorithmic latency compared with an offline S2S model trained with PIT.

The contributions of the paper are summarized as follows. 
\begin{itemize}
    \item We perform the first study on streaming end-to-end multi-talker speech recognition, and propose an RNN-T based model for this problem. We also demonstrate a strong recognition accuracy compared with an offline system.
    \item PIT is the mostly commonly used loss function for multi-talker speech processing, while an approach similar to HEAT has only been used in~\cite{tripathi2020end}. We present a rigid comparison of the two approaches in dealing with the label ambiguity problem, and show the superiority of HEAT over PIT in terms of computational efficiency and model accuracy for our problem.
    \item We propose two Unmixing architectures that are inspired from related works.
    % in CTC and S2S based end-to-end models as well as speech separation, we study two network architectures for the steaming multi-talker speech recognition problem. 
\end{itemize}

\section{Related Work}

There have been a few studies on S2S and joint CTC/attention models for end-to-end overlapped speech recognition~\cite{settle2018end, chang2019end, kanda2020serialized, kanda2020joint, kanda2020minimum}, however, these works are all in the category of offline condition. To the best of our knowledge, our work is the first study on {\it streaming} end-to-end overlapped ASR. The work that is most closely related to our work is RNN-T based approach for end-to-end overlapped ASR done by Tripathi et al.~\cite{tripathi2020end}. However, the authors also focus on the offline scenario in their work. In addition, the authors in~\cite{tripathi2020end} applied carefully designed auxiliary loss functions for signal reconstruction to train the RNN-T model, while in our work, we apply a single ASR loss function for model training, which simplifies the system development. Besides, the model architectures and loss functions are also different in this work.

\section{RNN-T}
\label{sec:rnnt}

RNN-T is a time-synchronous model for sequence transduction, which works naturally for end-to-end streaming speech recognition. Given an acoustic feature sequence $X=\{x_1, \cdots, x_T\}$ and its corresponding label sequence $Y=\{y_1, \cdots, y_U\}$, where $T$ is the length of the acoustic sequence, and $U$ is the length of the label sequence, RNN-T is trained to directly maximizing the conditional probability 
\begin{align}
\label{eq:prob}
P(Y \mid X) = \sum_{\tilde{Y} \in \mathcal{B}^{-1}(Y)}{P(\tilde{Y} \mid X)},
\end{align}
where $\tilde{Y}$ is a path that contains the blank token $\O$, and the function $\mathcal{B}$ denotes mapping the path to $Y$ by removing the blank tokens in $\tilde{Y}$. Essentially, the probability $P(Y \mid X)$ is calculated by summing over the probabilities of all the possible paths that can be mapped to the label sequence after the function $\mathcal{B}$. The probability can be efficiently computed by the forward-backward algorithm, which requires to compute the probability of each step~\cite{graves2012sequence}, i.e.,  
\begin{align}
P(k \mid x_{[1:t]}, y_{[1:u]}) = \frac{\exp \left( J(f_t^k + g_u^k)  \right)}{\sum_{k^\prime \in \bar{\mathcal{V}}}\exp \left(J(f_t^{k^\prime} + g_u^{k^\prime})  \right)},
\end{align}
where ${\bm f}_t$ and ${\bm g}_u$ are the output vectors from the audio encoder network and the label encoder network  followed by an affine transform at the time step $t$ and $u$ respectively, and $J(\cdot)$ denotes a nonlinear activation function followed by an affine transform. $\bar{\mathcal{V}}$ denotes the set of the vocabulary $\mathcal{V}$ with an additional blank token, i.e., $\bar{\mathcal{V}} = \mathcal{V} \cup \O$. Given the distribution of each timestep $(t, u)$, the sequence-level conditional probability Eq. \eqref{eq:prob} can be obtained by the forward-backward algorithm, where the forward variable is defined as
\begin{align*}
\alpha(t, u) & = \alpha(t-1, u) P(\O \mid  x_{[1:t-1]}, y_{[1:u]}) \\ 
 & + \alpha(t, u-1) P(y_u \mid x_{[1:t]}, y_{[1:u-1]}),
\end{align*}
with the initial condition $\alpha(1, 0) = 1$, while the backward variable can be defined similarly. The probability $P(Y \mid X)$ can be computed as
\begin{align}
P(Y \mid X) = \alpha(T, U) P(\O | x_{[1:T]}, y_{[1:U]}).
\end{align}
% The popular loss function to train RNN-T is the log-loss, which simply minimize the negative log-likelihood as:
RNN-T is trained by minimizing the negative log-likelihood as:
\begin{align}
\label{eq:nll}
\mathcal{L_{\text{rnnt}}} (Y, X) = -\log P(Y \mid X)
\end{align}

\begin{figure}[t]
\small
\centerline{\includegraphics[width=0.35\textwidth]{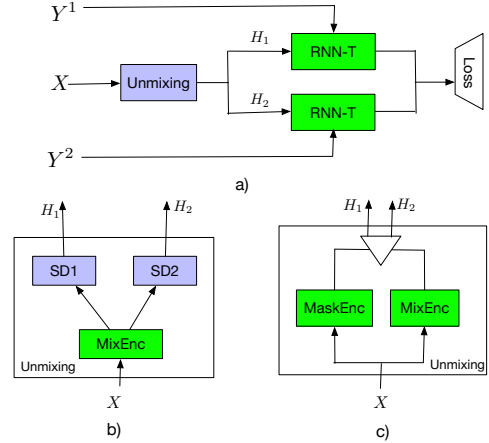}}
\vskip-3mm
\caption{a) RNN-T based SURT model. Here the RNN-T modules share the same model parameters. b) The speaker-differentiator (SD) based Unmixing module. In this case, SD1 and SD2 have different model parameters. c) Mask-based Unmixing module, where $\bigtriangledown$ denotes the masking operation (c.f. \ref{sec:mask}). }  
\label{fig:surt}
\vskip -5mm
\end{figure}

\section{Streaming Unmixing and Recognition Transducer}
In this work, we focus on the 2-speaker case for overlapped speech recognition, and the proposed SURT model is shown in Fig.\ref{fig:surt}-a).  We denote the overlapped acoustic sequence as $X$, and the label sequences are $Y^1$ and $Y^2$. Given the overlapped speech $X$, the Unmixing module extracts the speaker-dependent features representations, $H_1$ and $H_2$, which are then fed into the RNN-T module. The role of the Unmixing module is similar to speech separation, however, we do not apply any speech separation loss in training. Instead, the whole SURT model is trained end-to-end using a speech recognition loss as defined in section \ref{sec:loss}. In this section, we firstly discuss two network structures for the Unmixing module, before explaining the loss functions to train the models. 

% \begin{figure}[t]
% \small
% \centerline{\includegraphics[width=0.22\textwidth]{mask-rnnt.eps}}
% \caption{The mask-based network for the 2-speaker case. All model parameters are globally shared. In this figure, $\tilde{M} = \mathds{1} - M$. $\mathds{1}$ is tensor of the same shape as $M$, and each of its element is 1. }  
% \label{fig:mask}
% \vskip -5mm
% \end{figure}

\subsection{Speaker-Differentiator based Unmixing Model}
\label{sec:sd}

Inspired by~\cite{chang2019end}, we use two speaker-differentiator (SD) encoders to construct the Unmixing module as shown in Fig.\ref{fig:surt}-b). The speaker-dependent feature representations $H_1$ and $H_2$ are obtained as 
\begin{align*}
    \bar{X} = \text{MixEnc}(X), \hskip2mm H_1 = SD1 (\bar{X}), \hskip2mm H_2 = SD2(\bar{X}),
\end{align*}
where MixEnc is an encoder used to pre-process the overlapped speech signals; SD1 and SD2 are two difference encoders to generate the two feature sequences. While many different neural network encoders are applicable, we focus on convolution neural networks (CNNs) in this work as detailed in the experimental section.
% from the mixed audio signal. These two encoders have different model parameters. Following~\cite{chang2019end}, we use a shared mixture encoder to pre-process the mixture signals before feeding them to the SD encoders. The outputs of the two SD encoders are then fed into the shared RNN-T network to compute the loss. The network structure is shown in Figure~\ref{fig:sd} in the case of HEAT loss (cf. section~\ref{sec:heat}).

\subsection{Mask-based Unmixing Model}
\label{sec:mask}

Inspired by works in speech separation~\cite{wang2014training, yu2017permutation}, we define a mask-based Unmixing module as Fig.\ref{fig:surt}-c), in which, $H_1$ and $H_2$ are obtained as
\begin{align*}
    M &= \sigma(\text{MaskEnc}(X)), \hskip3mm \bar{X} = \text{MixEnc}(X), \\
    H_1 &= M * \bar{X}, \hskip1cm H_2 = (\mathds{1} - M) * \bar{X},
\end{align*}
where $\sigma$ denotes the Sigmoid function, and MaskEnc is the encoder to estimate the mask $M$; MixEnc is the pre-processing encoder as discussed before, and  $\mathds{1}$ is a tensor of the same shape as $M$, and each of its elements is 1; $*$ denotes element-wise multiplication.
% $X_1$ and $X_2$ are then fed into the shared RNN-T network to compute the loss. The network structure is shown in Figure~\ref{fig:mask} in the case of HEAT loss (cf. section~\ref{sec:heat})..

\subsection{Loss Functions}
\label{sec:loss}

% We denote the two feature representations as $H_1$ and $H_2$ as the input sequences to the RNN-T module of a SURT model. For the SD-based model, $H_1$ and $H_2$ are the output hidden vectors from the two SD encoders, while for the mask-based model, they correspond to $X_1$ and $X_2$, respectively. 
For model training, we study two loss functions, i.e., Permutation Invariant Training (PIT)~\cite{yu2017permutation} and Heuristic Error Assignment Training (HEAT). %~\cite{tripathi2020end}.

\subsubsection{Permutation Invariant Training}
\label{sec:pit}

PIT~\cite{yu2017permutation} has been widely used for speech separation and multi-talker speech recognition due to its simplicity and superior performance. The key problem in overlapped speech separation and recognition, as argued in~\cite{yu2017permutation}, is the label ambiguity issue, i.e., it is unclear if the feature representation $H_1$ corresponds to $Y^1$ or $Y^2$. To address this problem, PIT considers all the possible error assignments when computing the loss, and hence, it is {\it invariant} to the label permutations. For the 2-speaker case studied in this work, the PIT loss can be expressed as: 
\begin{align}
\mathcal{L}_{\text{pit}}(X, Y^1, Y^2) = &\min(\mathcal{L}_{\text{rnnt}}(Y^1, H_1) + \mathcal{L}_{\text{rnnt}}(Y^2, H_2),  \nonumber \\
& \mathcal{L}_{\text{rnnt}}(Y^2, H_1) + \mathcal{L}_{\text{rnnt}}(Y^1, H_2)).
\end{align}
While being simple and effective, PIT also has drawbacks. In particular, it is not very scalable to the number of speakers in the mixed signal. For the $S-$speaker case, the total number of permutations is $S!$, which will require to compute the RNN-T loss $S!$ times in the framework of SURT.
We could use Hungarian algorithm \cite{kuhn1955hungarian} to reduce the computation from $O(S!)$ to $O(S^3)$, but it is still clearly not affordable due to the high computational and memory cost of the RNN-T loss. 

\begin{table}[t]\centering
\caption{The 2D CNN structure used in the MixEnc and MaskEnc used in SD- and Mask-based Unmixing modules. The shape for conv2d operator is (input\_channel, output\_channel, kernel\_width, kernel\_height).}
\label{tab:cnn}
\footnotesize
\vskip-2mm
\begin{tabular}{ccc}
\hline 

\hline
 Type & Depth & Shape \\ \hline
Conv2D & 4 & \( \left[ \begin{array}{c} 
{\tt conv2d}(3, 64, 3, 3) \\ 
{\tt conv2d}(64, 64, 3, 3) \\
{\tt Maxpool}(3, 1) \\
{\tt conv2d}(64, 128, 3, 3) \\
{\tt Maxpool}(3, 1) \\
{\tt conv2d}(128, 128, 3, 3) \\
{\tt Maxpool}(3, 1) \\
{\tt Linear} \end{array} \right] \) \\  \hline

\hline
\end{tabular}
\vskip-4mm
\end{table}

\subsubsection{Heuristic Error Assignment Training}
\label{sec:heat}

%In this work, we compare PIT with Heuristic Error Assignment Training (HEAT), which is computationally cheaper and memory-wise more efficient compared with PIT. 
Different from PIT, HEAT only picks one possible error assignment based on some heuristic information that can disambiguate the labels. %We argue that the permutation problem only happens for fully overlapped speech signals. %in the chunk-wise condition, in which the two segments from the two speakers are fully overlapped. 
%In the real scenario as well as in our experimental setup, the two speaker utterances are mostly partially overlapped at the utterance-level fashion. We believe that the model can be trained to rely on the information from the single-speaker region to track the corresponding speaker, and differentiate the feature representations from the competing speaker. Thus in HEAT, we only pick an error assignment between the feature representations $(H_1, H_2)$ and the labels $(Y^1, Y^2)$ based on some heuristic.
In this work we particularly use the heuristic to disambiguate the labels based on the start times that they were spoken, e.g.,
\begin{align}
\label{eq:heat}
\mathcal{L}_{\text{heat}}(X, Y^1, Y^2) = \mathcal{L}_{\text{rnnt}}(Y^1, H_1) + \mathcal{L}_{\text{rnnt}}(Y^2, H_2),
\end{align}
where $Y^1$ always refers to the utterance that was spoken first in our setting. %In~\cite{tripathi2020end}, the authors defined the mapping between $(H_1, H_2)$ and $(Y^1, Y^2)$ by applying the time masking using the oracle time boundaries of $Y^1$ and $Y^2$. However, in HEAT, we do not need to know the time information, but the information which utterance was spoken first. Since we do not reply on the time information, we do not use the auxiliary masking L2 loss and the embedding loss as in~\cite{tripathi2020end}, which makes our loss function much simpler.
Similar approach has been used in~\cite{tripathi2020end}, and the authors also tried other heuristic information such as the time boundaries which were used to mask the encoder embedding vectors and define the mapping between $(H_1, H_2)$ and $(Y^1, Y^2)$. They also introduced auxiliary loss functions, while in our work, we prefer Eq.~\eqref{eq:heat} for simplicity. With HEAT, the model will be trained to produce the hidden representations $H_1$ that match the label sequence $Y^1$. %In~\cite{tripathi2020end}, the authors masked the elements that are outside of the oracle time boundaries of $Y^1$ to be zero, while in our approach, the model is trained to emit blank tokens $\O$ for the corresponding time steps. During inference, our model can be free from the mask estimation. 
Note that, it does not make any difference if we swap $H_1$ and $H_2$, as before model training, the model parameters do not have any label correspondence yet. However, once the mapping function is chosen, we have to fix it during model training. Compared with PIT, HEAT is more scalable and memory efficient, as it only evaluates the RNN-T loss $S$ times for the $S$-speaker case. % we only need to evaluate the RNN-T loss function $S$ times, instead of $S!$ times as in PIT. 
% \begin{table}[t]\centering
% \caption{Mask-based model architecture. The structure of the Mask encoder is the same as the Mixture encoder in the SD-based network, except that the top layer is a Sigmoid activation function.}
% \label{tab:mask}
% \footnotesize
% \vskip-2mm
% \begin{tabular}{c|ccc}
% \hline 

% \hline
% Module & Type & Depth & Shape \\ \hline
% Mask & Conv2D & 4 & \( \left[ \begin{array}{c} 
% {\tt conv2d}(3, 64, 3, 3) \\ 
% {\tt conv2d}(64, 64, 3, 3) \\
% {\tt Maxpool}(3, 1) \\
% {\tt conv2d}(64, 128, 3, 3) \\
% {\tt Maxpool}(3, 1) \\
% {\tt conv2d}(128, 128, 3, 3) \\
% {\tt Maxpool}(3, 1) \\
% {\tt Linear} \\
% {\tt Sigmoid} \end{array} \right] \) \\  \hline
% RNNT-A & LSTM & 6 & (771, 1024) \\
% RNNT-L & LSTM & 2 & (1024, 1024) \\ \hline

% \hline
% \end{tabular}
% \end{table}

\begin{figure}[t]
\small
\centerline{\includegraphics[width=0.45\textwidth]{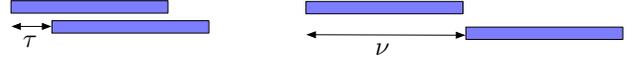}}
\caption{Overlapped speech simulation. $\tau$ refers to the minimum delay, and $\nu$ refers to the maximum delay, which is the length of the first utterance.  }  
\label{fig:data}
\vskip -5mm
\end{figure}

\section{Experiments and Results}
\label{sec:exp}

\subsection{Dataset}
\label{sec:data}

Our experiments were performed on the simulated LibriSpeechMix dataset~\cite{kanda2020serialized}, which is derived from the 1,000 hour LibriSpeech corpus~\cite{panayotov2015librispeech} by simulating the overlapped audio segments. We used the same protocol to simulate the training and evaluation data as in~\cite{kanda2020serialized}. The source code to reproduce our evaluation data is publicly available\footnote{\scriptsize \url{https://github.com/NaoyukiKanda/LibriSpeechMix}}.  To generate the simulated training data, for each utterance in the original LibriSpeech {\tt train\_960} set, we randomly pick another utterance from a different speaker, and mix the latter with the previous one with a random delay sampled from $[\tau, \nu]$, in which $\tau$ and $\nu$ are the minimum and maximum delay in seconds respectively, as shown in Figure~\ref{fig:data}. $\nu$ is always the same as the length of the first utterance, and we evaluate two different values of $\tau$ in our experiments, i.e., $\tau = 0$ and $\tau = 0.5$. We used the same approach to generate the {\tt dev-clean} and {\tt test-clean} datasets. The number of mixed audio is the same as the number of utterances in the original LibriSpeech dataset. For both training and evaluation data, each utterance only has 2 speakers after simulation.

\subsection{Experimental Setup}
\label{sec:setup}

In our experiments, we used the magnitude of the 257-dimensional short-time Fourier transform (STFT) as raw input features, which are sampled as the 10 milliseconds frame rate. The features were then spliced by a context window of 3 and downsampled by a factor of 3, results in 771-dimensional features at the frame rate of 30 milliseconds. We then reshaped the feature sequences to have 3 input channels. We used 4,000 word-pieces as the output tokens for RNN-T, which are generated by byte-pair encoding (BPE)~\cite{sennrich2015neural}. We set the dropout ratio as 0.2 for LSTM~\cite{hochreiter1997long} layers, and applied one layer of time-reduction to further reduce the input sequence length by the factor of 2~\cite{graves2012hierarchical, chan2016listen, lu2016segmental}. We also applied speed perturbation for data augmentation with perturbation ratios as 0.9 and 1.1~\cite{ko2015audio}. 

We used 4-layer 2D CNN encoder for MixEnc in the Unmxing module of the SD-based SURT model. The detailed configuration is shown in Table~\ref{tab:cnn}. We used a 2-layer unidirectional LSTM with 1024 hidden units for SD1 encoder, SD2 encoder, the audio encoder and the label encoder of RNN-T. For the  mask-based SURT model, we used the same CNN encoder as in Table~\ref{tab:cnn} for MixEnc and MaskEnc for the Umixing module. The label encoder of RNN-T is a 2-layer unidirectional LSTM with 1024 hidden units as in the SD-based model, and the audio encoder is a 6-layer unidirectional LSTM with 1024 hiddent unit. The total number of model parameters is around 80 million (M) for both model architectures, and the algorithmic latency for both types of model is 5 frames, corresponding to 150 milliseconds, which is incurred by the convolution module. In our experiments, the models were trained using Adam optimizer~\cite{kingma2014adam} with the intial learning rate as $4\times 10^{-4}$, and halved the learning rate every 40,000 updates.  We used data parallelism across 16 GPUs, and the mini-batch size for each GPU is 5,000 frames for both model architectures. During evaluation, the model produces two transcriptions in the 2-speaker case. For scoring, we follow the same protocol as in~\cite{tripathi2020end,kanda2020serialized} by choosing the label permutation yielding the lowest word error rate (WER). 

\begin{table}[t]\centering
\caption{Results of SD-based SURT models trained with PIT and HEAT. We evaluate two data simulation conditions, i.e., $\tau = [0, 0.5]$. }
\label{tab:results}
\footnotesize
\vskip-2mm
\begin{tabular}{l|l|l|cc}
\hline 

\hline
Model & Training Data & Loss & \multicolumn{2}{c}{dev-clean}  \\
&  &          & $\tau=0$  & $\tau=0.5$   \\ \hline
& $\tau = 0.5$ & PIT & 12.0 &  11.3  \\ 
SD &   & HEAT & 11.8  & 10.9  \\ \cline{2-5}
& $\tau = 0$  & PIT & 13.1   & 11.8  \\
&& HEAT & 12.5   & 11.2    \\
\hline

\hline
\end{tabular}
\vskip-2mm
\end{table}

\begin{table}[t]\centering
\caption{Comparison of different network structures. }
\label{tab:sd_mask}
\footnotesize
\vskip-2mm
\begin{tabular}{l|l|cc}
\hline 

\hline
  Model & Loss & \multicolumn{2}{c}{dev-clean}  \\
  &          & $\tau=0$ & $\tau=0.5$ \\ \hline
 SD & PIT & 12.0 &  11.3  \\ 
 & HEAT & 11.8 & 10.9  \\ \hline
Mask w/o MixEnc & PIT &  14.1  & 13.8  \\
& HEAT & 13.4  & 12.3    \\ \hline
Mask w/ MixEnc & HEAT & 10.1  & 9.5   \\ \hline

\hline
\end{tabular}
\vskip-4mm
\end{table}

\subsection{Results}
\label{sec:results}

Table~\ref{tab:results} shows the WER results of the SD-based model. In particular, we evaluated two conditions when generating the mixed speech signals, i.e., $\tau = [0, 0.5]$, for both training and evaluation data. From the results in Table~\ref{tab:results}, we observe that using the training data with the minimum delay $\tau = 0.5$, the model achieved consistent lower WERs in both evaluation conditions compared with the model trained with data of $\tau = 0$. Our interpretation is that the starting region of the speech signal that has no overlap can provide a strong cue for the model to track the first speaker and disentangle the overlapped signals. This information also makes the recognition task easier, as we observe that the model can achieve consistent lower WER for the evaluation condition $\tau = 0.5$ compared with the evaluation condition of $\tau = 0$. 

\begin{figure}[t]
\small
\centerline{\includegraphics[width=0.35\textwidth]{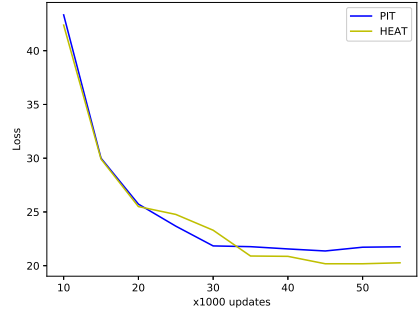}}
\caption{Comparison of HEAT and PIT loss functions in terms of validation loss values. The two approaches can yield similar convergence speed, while HEAT can reach lower validation loss compared with PIT.}  
\label{fig:loss}
\vskip -2mm
\end{figure}

The results also shown that HEAT can achieve lower WERs compared with PIT. To further understand the behaviors of the two loss functions, we plot the convergence curves of the models trained with PIT and HEAT in Figure~\ref{fig:loss}. The horizontal axis indicates the validation loss values, while the vertical axis represents the number of model updates. In this comparison, we used exactly the same experimental setting for model training. The figure shows that the two approaches can result in very similar convergence speed, and HEAT can reach to a lower validation loss. As discussed before, HEAT is also faster than PIT, and we can use a larger mini-batch size as HEAT requires less memory.

\begin{table}[t]\centering
\caption{Comparison with PIT-S2S model. Latency refers to the algorithmic latency in terms of millisecond.}
\label{tab:comp}
\footnotesize
\vskip-2mm
\begin{tabular}{l|lllc}
\hline 

\hline
Training data & Model & Size & Latency & $\tau$ = 0  \\
&  &        &    & test-clean   \\ \hline
% $\tau = 0.5$ & SURT & 80M & 120 &  11.75 & 11.70 \\ 
$\tau = 0.5$ & SURT & 81M & 150  & 10.8 \\ 
&PIT-S2S~\cite{kanda2020serialized} & 160.7M & $\infty$  & 11.1   \\
\hline

\hline
\end{tabular}
\vskip-5mm
\end{table}

Table~\ref{tab:sd_mask} compares the SD-based model with the Mask-based model with (w/) or without (w/o) the MixEnc encoder, and the results show that the Mask-based model achieved much lower WERs. Finally, Table~\ref{tab:comp} compares the proposed Mask-based SURT model (w/ MixEnc) with an offline LSTM-based S2S model trained with PIT~\cite{kanda2020serialized}. SURT achieved comparable results with half of the number of model parameters and with a very low latency constraint. % SURT only falls slightly behind the offline PIT-S2S model in terms of the WER. It demonstrates that SURT points out a promising research direction for streaming end-to-end overlapped speech recognition. 

\section{Conclusions}
\label{sec:conc}

Overlapped speech recognition remains a challenging problem in the speech research community. While all the existing end-to-end approaches  tackling this problem work in the offline condition, we proposed Streaming Unmixing and Recognition Transducer (SURT) for end-to-end multi-talker speech recognition, which can meet various latency constraints. In this work, SURT relies on RNN-T as the backbone, while other types of streaming transducers such as Transformer Transducers \cite{zhang2020transformer, chen2020developing} are also applicable. We investigated two different model architectures, and two different loss functions for the proposed SURT model. Based on experiments using the LibrispeechMix dataset, we achieved strong recognition accuracy with very low latency and a much smaller model compared with an offline PIT-S2S model.

\bibliographystyle{IEEEtran}

\bibliography{bibtex}

\end{document}